\documentstyle[sprocl]{article}

\input psfig.sty

\def\be{\begin{equation}}
\def\ee{\end{equation}}
\def\bea{\begin{eqnarray}}
\def\eea{\end{eqnarray}}

\def\li11{$^{11}Li$}
\def\o18{$^{18}O$}

\begin{document}

\hspace{3in} {UConn-40870-0030}

\title{{\Large \underline{{\bf Nuclear Molecular Halo:}}} \\ 
   Threshold Effect or Soft Dipole in $^{11}Li$? \footnote{Work 
Supported by USDOE Grant No. DE-FG02-94ER40870.}}

\author{Moshe Gai
\footnote{Invited Talk, XXII Symposium on 
Nucl. Phys., Jan. 6-9, 1999, Oaxtepec, Mexico.}}

\address{Dept. of Physics, U46, University of Connecticut,
   2152 Hillside Rd., \\ Storrs, CT 06269-3046, USA; 
   gai@uconnvm.uconn.edu; http://www.phys.uconn.edu}


\maketitle

\abstracts{The observation of large E1 strength near threshold in the 
electromagnetic dissociation of $^{11}Li$ poses a fundamental question:  
Is the large E1 strength due to the threshold or is it due to a low lying 
E1 state? Such molecular cluster states were observed in $^{18}O$ 
and in several nuclei near the drip line. 
We discuss the nature of the "threshold effect" as well as review the 
situation in Molecular (and Particle Physics) where such Molecular States 
are observed near the dissociation limit.  We suggest that the situation in 
$^{11}Li$ is reminiscent of the argon-benzene 
molecule where the argon atom is 
loosely bound by a polarization (van der Waals) mechanism and thus leads to 
a very extended object lying near the dissociation limit. Such states are 
also suggested to dominate the structure of mesons [$a_0(980),\ f_0(975)$] 
and baryons [$\Lambda(1405)$] with proposed 
Kaon molecular structure (e.g. by 
Dalitz) near threshold. The inspection of 
such states throughout Physics allows
us to gain insight into this phenomenon and suggest that a new collective 
Molecular Dipole Degree of Freedom plays a major role in the 
structure of hadrons (halo nuclei, mesons 
and baryons), and that quantitative 
tools such as the E1 Molecular Sum Rule are useful for elucidating 
the nature of the observed low lying E1 strength in halo nuclei.}

\section{Introduction: Dipole (E1) Strength in \li11}

A measurement of the electromagnetic dissosiation of \li11 was 
performed at GSI \cite{GSI} from which the electric dipole (E1) 
strength shown in Fig. 1 was extracted.

\centerline{\psfig{figure=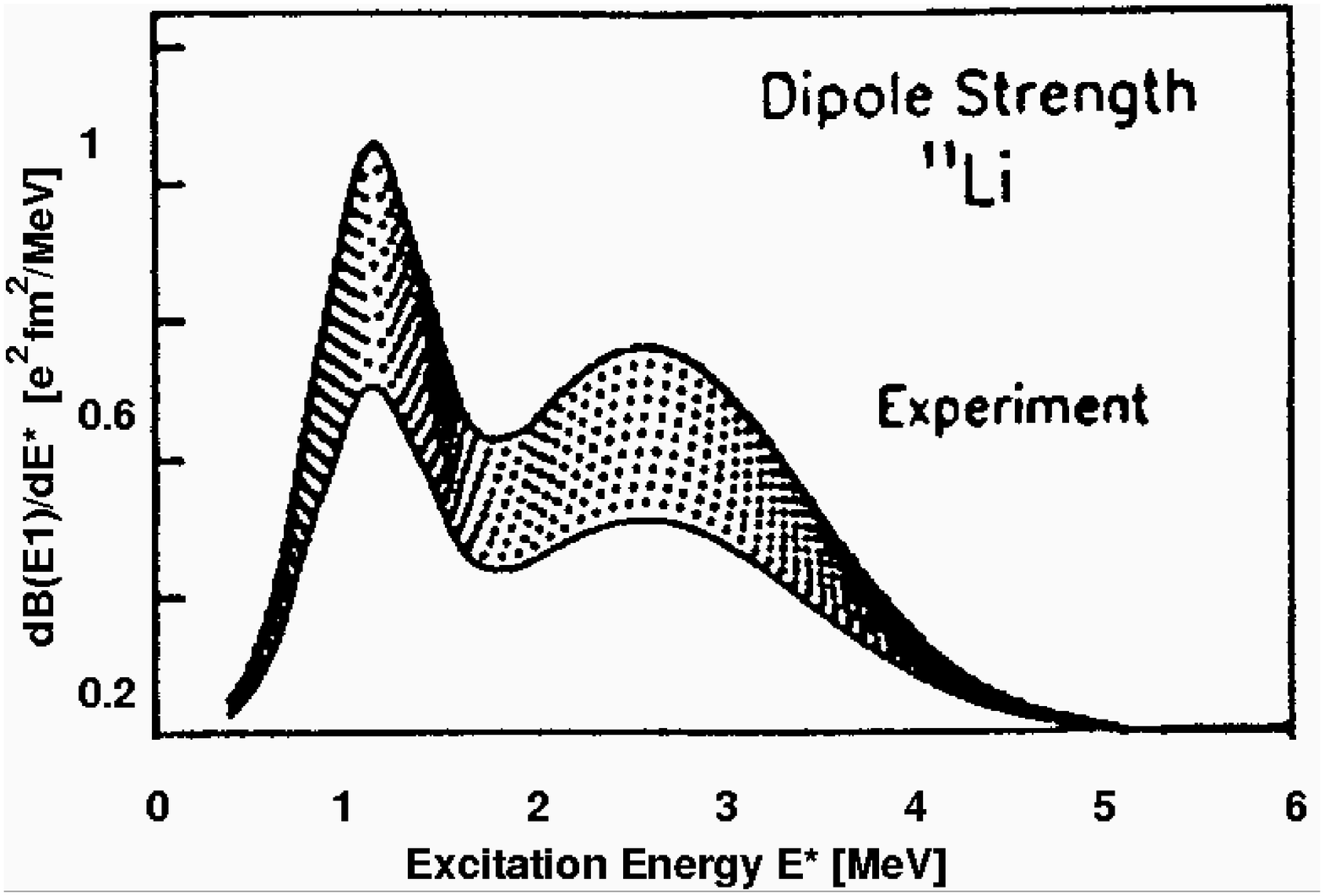,height=3.5in}}

\underline{Fig. 1:}  Electromagnetic dipole (E1) strength of \li11 
   as measured by the GSI collaboration \cite{GSI}.

This measurement poses a fundamental question: Is the large E1 
strength just above 1 MeV in \li11, see Fig. 1, due to the kinematics  
(a "threshold effect") or is it due to the structure of \li11 (a "low 
lying dipole state")? In this talk we examine these questions, we 
define and test the concept of a "threshold effect", as well as 
examine molecular structure in molecular and particle physics 
near threshold. We show that quantitative tools exist that allow 
us to examine this question in details.

The possibility of such a low lying dipole state \cite{Ik92,Ha95} 
was suggested by the RIKEN data \cite{Ko97} on proton 
scattering off \li11 where a peak was observed at approximately 
1.3 MeV.  The poor resolution of this experiment however does not 
permit a determination of the intrinsic width of such a state or to 
disintangle it from an underlying broad background. The possibility 
of a low lying dipole state is also given credence by the 
previous RIKEN data \cite{Kob92} on the pion double charge exchange 
off $^{11}B$ where an $\ell = 1$ state at approximately 1.2 MeV 
in \li11 was suggested. The pion data were given (some) credence 
recently \cite{Gor98}.  However the MSU group \cite{Ka97} has 
recently proposed a "nuclear shakeoff mecahnism" that explains this 
bump without invoking a low lying dipole state in \li11. In such 
a mechanism the proton (as well as the photon) imparts its momentum 
to the $^9Li$ core, and thus "shaking off" the two neutrons. Such a 
mecahnism leads to a strength with a maximum at 1.3 MeV and a 
high energy tail, as observed \cite{Ko97} in the low resolution 
experiment at RIKEN. The MSU group states \cite{Ka97}: "In 
conclusion, there does not seem to be any compelling evidence 
from the proton scattering experiments of Korsheninnikov {\em et 
al.} \cite{Ko97} for a 1.3 MeV excited level in \li11."

To examine the "nuclear shakeoff mechanism" one obviously needs a 
probe that is strongly surface interacting. In this case the momentum 
could not be solely transferred to the $^9Li$ core and necessarily 
involves the "halo" neutrons at the surface. Such a probe is 
the pion (as well as other probes) and we conclude that the double 
pion scattering data of the RIKEN group \cite{Kob92} pose some 
difficulty to the "nuclear shakeoff mechanism".

\section{Threshold Effect: Photodisintegration of the Deutron and $^8Li$}

The photodisintegration of the deutron shown in Fig. 2, provides a 
vivid example of a "threshold effect".

\centerline{\psfig{figure=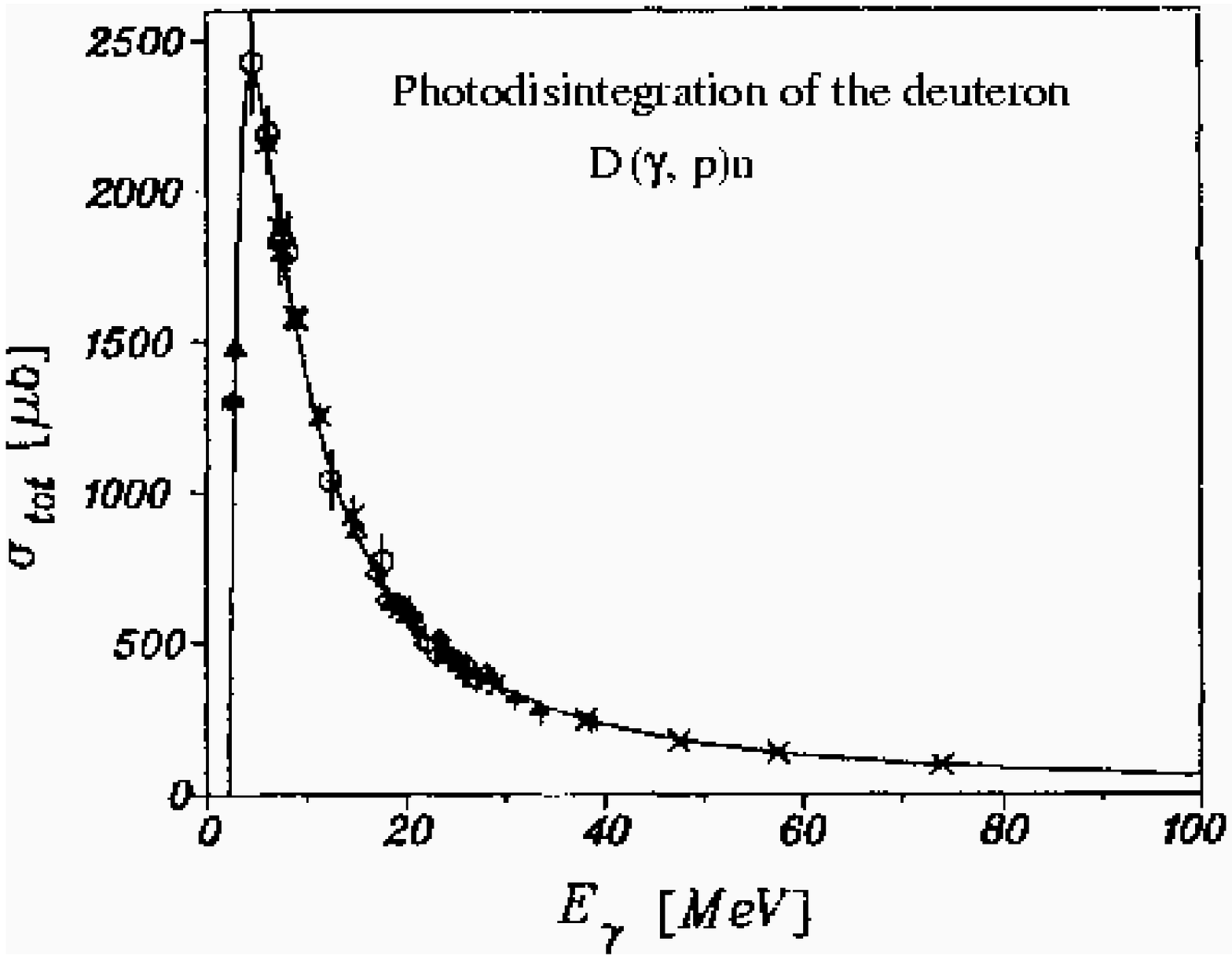,height=3.5in}}

\begin{center}
\underline{Fig. 2:}  The photodisintergation of the deutron.
\end{center}
Namely, the peak shown in Fig. 2 
at approximately 4.4 MeV does not corresponds 
to a state in the proton-neutron system and arise from the kinematics 
as we discuss below.

The photo nuclear cross section is derived using standard 
notation from detailed balance:
\   \\

\begin{tabbing} 

$\sigma (\gamma,n)$  \= $= \ {{(2J_1+1)(2J_2+1)} \over {2(2J_3+1)}} \ 
 {{k^2} \over {k_\gamma ^2}} \ \sigma (n,\gamma)$ \\
  \> $= \ {{(2J_1+1)(2J_2+1)} \over {2(2J_3+1)}} \ 2\mu c^2 \ 
  {E \over E_\gamma ^2} \ \sigma (n,\gamma)$ \\ 
  \> $= \ {{(2J_1+1)(2J_2+1)} \over {2(2J_3+1)}} \ 2\mu c^2 \ 
   \sigma (n,\gamma) \ {{E_\gamma - Q} \over E_\gamma ^2} $
  \hspace{1in} (equ. 1)
\  \\
\end{tabbing}
where the factor of 2 in the denominator arises from the two 
polarization states of a real transverse photon and Q = 2.223 MeV, 
is the one neutron separation energy in deuterium.
The kinematical factor ${{E_\gamma - Q} \over E_\gamma ^2} $ produces 
a peak at 2Q in the $\sigma (\gamma,n)$ cross section even in the 
absence of a peak (e.g. a state) in the $\sigma (n,\gamma)$ cross 
section. And we conclude that the peak at 4.4 MeV (= 2Q) in the 
photodisintegration of the deutron is solely due to that kinematical 
factor and we define it as a "threshold effect".  But we note that 
in the electromagnetic dissociation of \li11 one observes a peak 
at approximately 1.2 MeV, see Fig. 1, which is four times the two neutron 
separation in \li11 (Q = 300 keV), and hence this peak 
in \li11 can not 
arise from the above kinematical factor (of equ. 1) alone.

The capture of slow neutrons by nuclei is well understood 
\cite{Blatt,Sh58,Ber61,Bor82} and can be expanded in terms of the 
neutron velocity ($v$):

\begin{tabbing} 

$\sigma(n,\gamma)$ \= = $(\sigma E^{1/2})_0 \ [E^{-1/2} \ + \ \alpha \ 
   + \ \gamma E^{1/2} \ + \ ...] $\\
   \> = $(\sigma v)_0 \ [{1 \over v} \ + \ \alpha \ + \ 
     \gamma v \ + \ ...]$
   \hspace{1.4in} (equ. 2)
\end{tabbing} 
and for thermal neutrons $v_0$ = 2,200 m/s ($\beta_0 = 
7.3 \times 10^{-6}$) and $\sigma_0$ = 333 mb for the 
$p(n,\gamma)d$ reaction, thus $(\sigma v)_0$ 
= 2.5 $\mu$bc. For spin zero particles \cite{Sh58} we have $\alpha= 
{m \over \pi \hbar ^2} {A \over {A+1}} ^2$. The interaction of 
slow neutrons (up to a few hundred keV) is dominated by s-waves
with a large wave length, and is therefore dependent mainly 
on the time spent near the target giving rise to 
the well known 1/$v$ dependence of the cross section. At higher 
energies, p-waves dominate and the $\alpha$ terms 
is most important, and indeed the 
photodisintegration of the deuteron is given by that term (for 
$\alpha = 0.8$) as shown in Fig. 3. In the same figure we also show 
the (p-wave) continuum E1 calculated by Bethe 
and Longmire \cite{Bethe} (which clearly does 
not arise from an E1 dipole state at 4.4 MeV in the deuteron). We 
conclude that the peak at 2Q = 4.4 MeV in the photodisintegration 
of the deutron is a manifestation of a "threshold effect".

\centerline{\psfig{figure=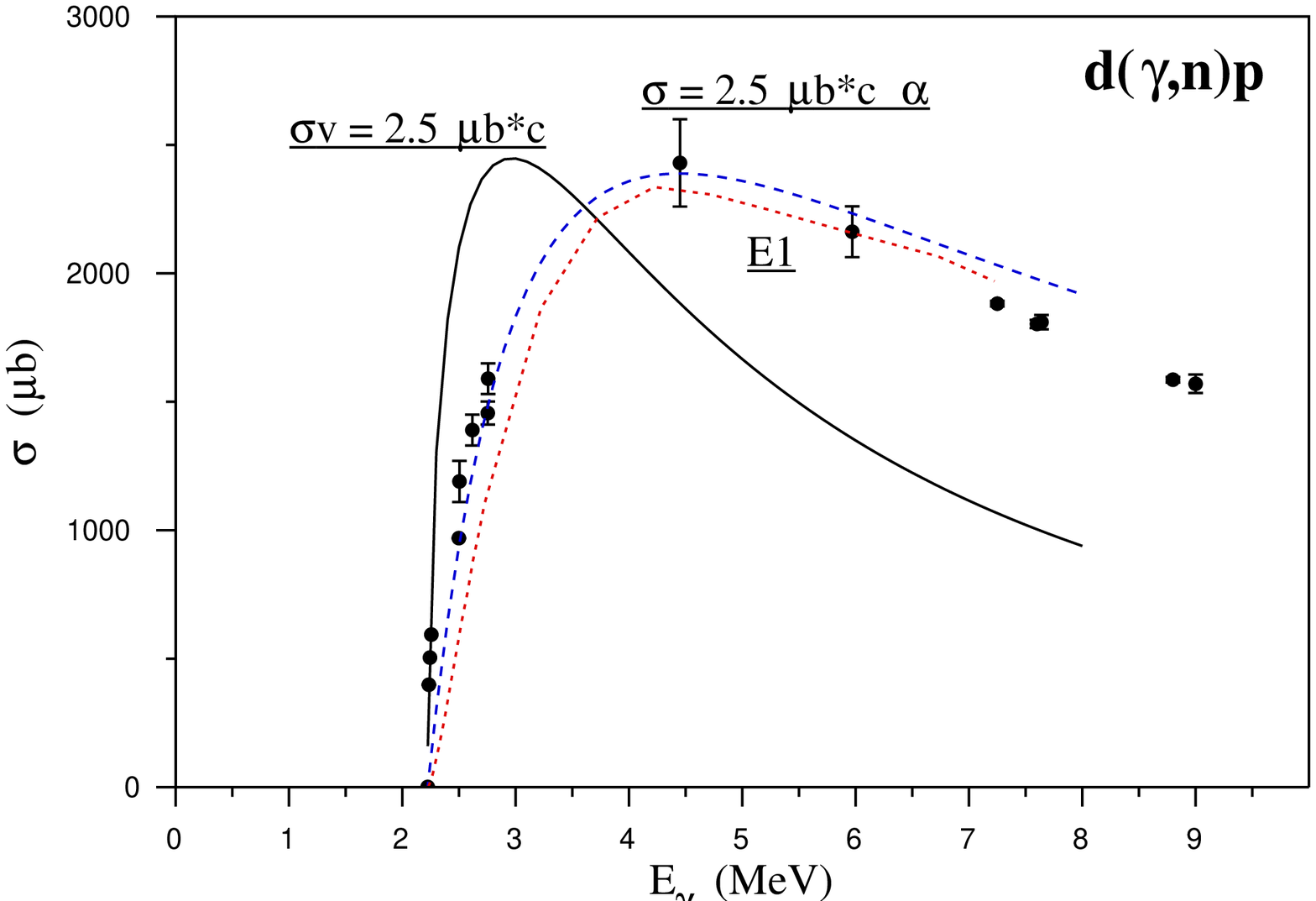,height=2.2in}}

\underline{Fig. 3:}  The photodisintergation of the deutron 
  as described by the second term ($\alpha$) of equ. 2.

The photodisintegration of $^8Li$ represnets yet another 
good example of a threshold effect.  In Fig. 4 we show these 
data as deduced from the direct capture of neutrons on $^7Li$ 
\cite{He98}. The interaction of the low energy neutrons is indeed 
dominated by s-waves \cite{He98} and the cross section of the 
$^7Li(n,\gamma)^8Li$ follwos the 1/$v$ law, as shown in Fig. 4.
The photodisintegration cross section is given by the first term 
of equ. 2, with $(\sigma v)_0 \ = \ 7.3 \times 10^{-6} \times 40 \ 
= \ 0.29 \ \mu b c$. Note the observation of (an interfering) 
$3^+$ state on top of a threshold effect.

\centerline{\psfig{figure=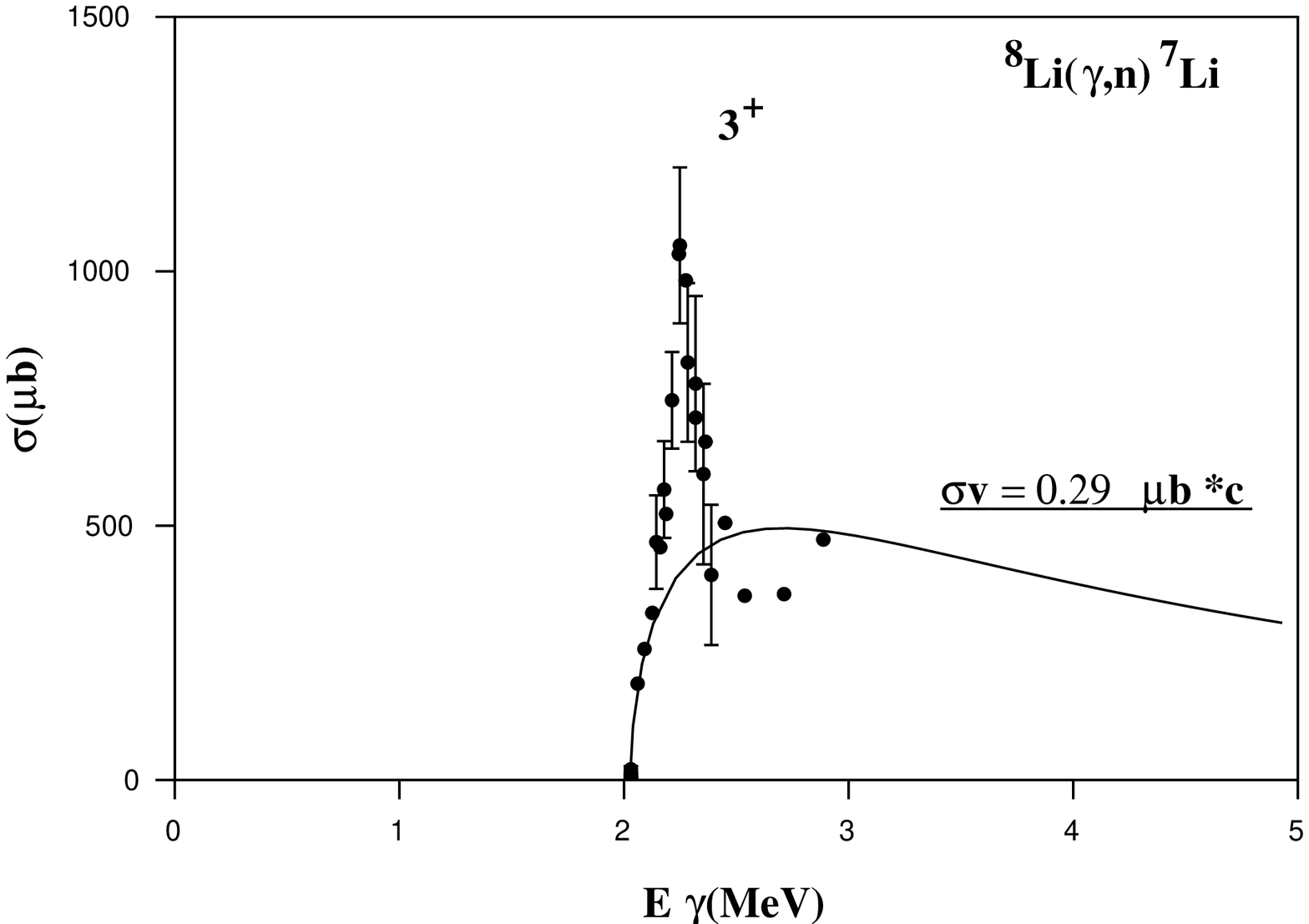,height=2.2in}}

\underline{Fig. 4:}  The photodisintergation of the $^8Li$, 
derived from direct capture data \cite{He98}, as described by 
the first term of equ. 2-- the 1/$v$ law.

\section{The Photodisintegration of $^{11}Li$}

The photodisintegration of $^{11}Li$ below 1.0 MeV yields 
two low energy neutrons with velocities that may differ.  But 
the two neutrons are in fact observed to emerge with almost 
identical energies (see Fig. 10 of \cite{GSI}), as one may 
expect for halo neutrons. Hence we make the 
assumption that the two neutrons are emitted with 
the same velocity (but not implying a physical di-neutron object)
and we use this velocity in the paramaterization of equ. 2. For these  
low energy neutrons (approx. 300 keV \cite{GSI}) 
we also expect the 1/$v$ law, as discussed above.
In contrast the GSI data can not be described by the 1/$v$ 
law, or any of the terms of equ. 2, as shown in Fig. 5. In 
particular the third term in equ. 2 (for $\gamma \ = \ 0.6$) 
does not yield a peak at 1.2 MeV.
However the shape of the spectrum is sufficiently uncertain that we 
can not rule out a "threshold effect" and this analysis thus 
calls for more accurate data on the shape of the spectrum,  
so as to test the validity of "threshold effect". 
Note that for a single step capture of two neutrons 
with two different velocities, 
the low velocity neutron tends to push the shape to lower 
energies, considerably below 4Q = 1.2 MeV, as discussed above.

\centerline{\psfig{figure=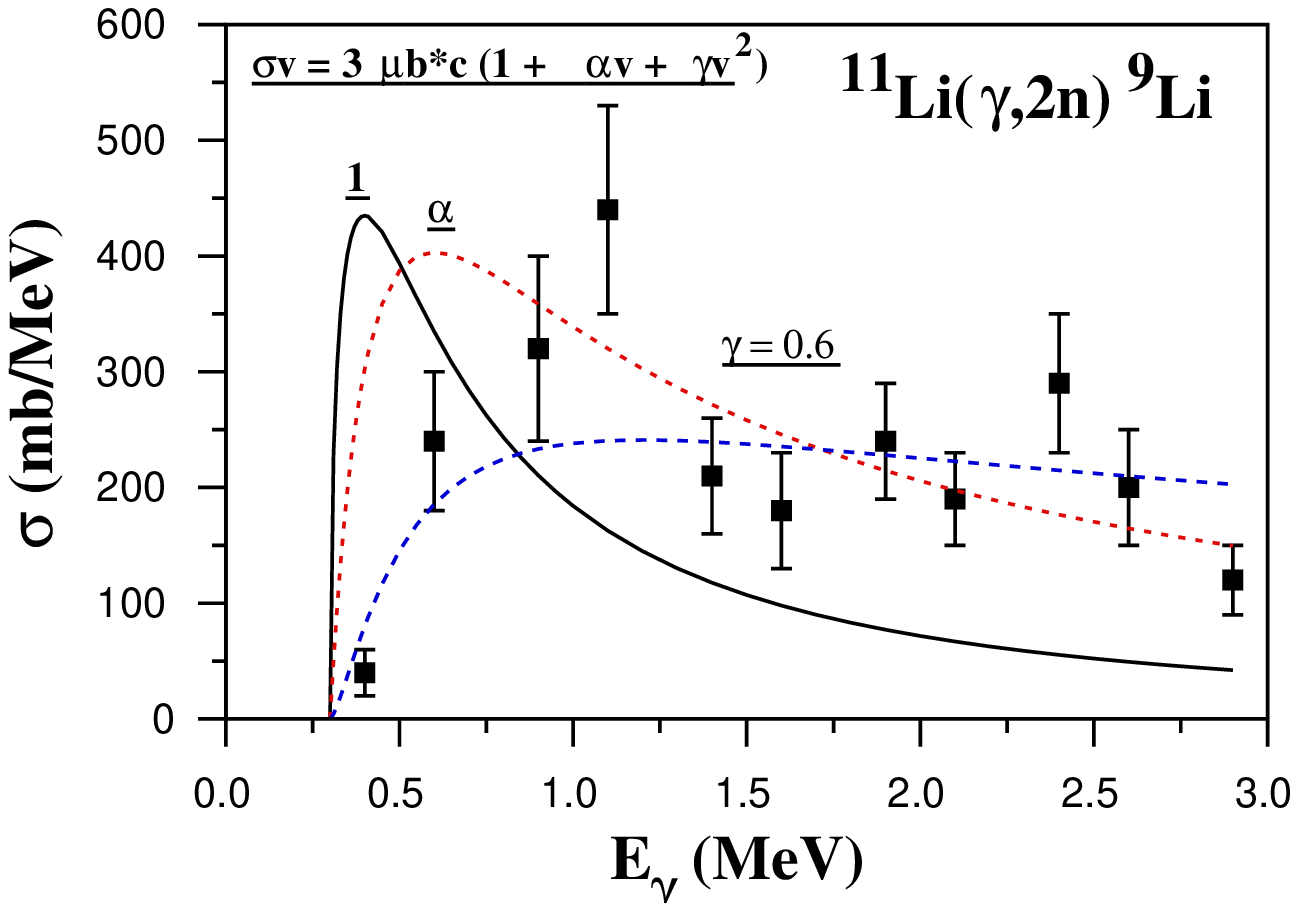,height=2.3in}}

\underline{Fig. 5:}  Electromagnetic dipole (E1) strength of \li11 
   \cite{GSI} and attempts to describe it by the terms of equ. 2.

\section{Molecular States in Molecular and Particle Physics}

The ubquitous occureence of molecular states {\bf near threshold} in Physics 
may indeed allow for insight into the structure of \li11 and other 
such "halo" nuclei. In Fig. 6 we show characteristic dimensions of the 
Ar-benzen molecule. The argon atom is 
losely bound to the (tightly bound) 
benzen molecule by a van der Waalls polarization and thus this 
molecular state lies  
close to the dissociation limit. We note that the relative dimension 
and indeed the very polarization phenomena are reminscent of 
the structure of \li11 where the argon atom creates a "halo" around 
the benzen molecule.

\centerline{\psfig{figure=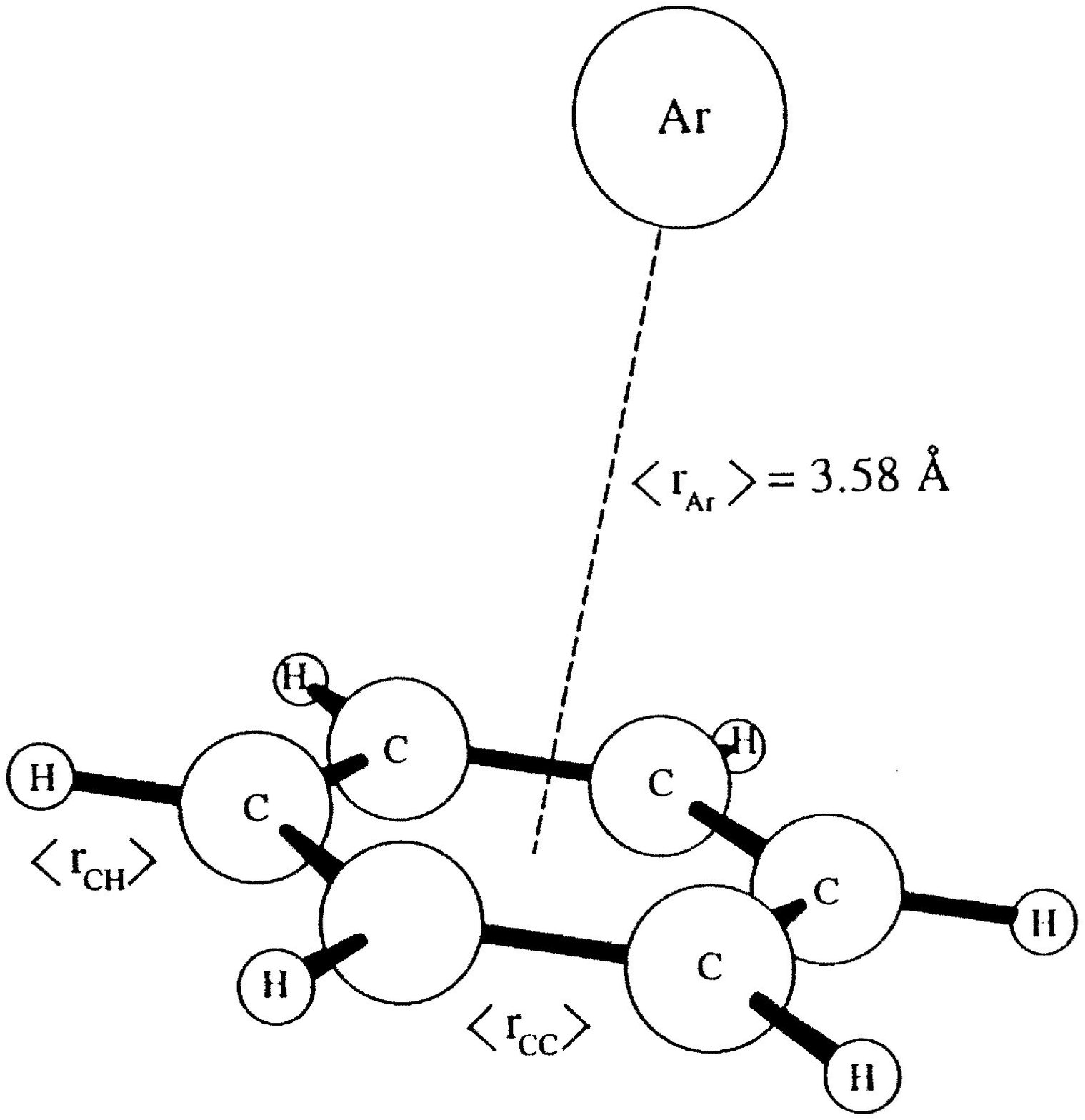,height=2in}}

\underline{Fig. 6:}  Characterstics dimensions of the Ar-benzen 
  molecule, adopted from Iachello and Levine \cite{Levine}.

Indeed the structure of baryons and mesons near threshold was 
suggested to be governed by a molecular degree of freedom. The 
$\Lambda(1405)$ that lies close to the N + Kbar threshold (1435 MeV) 
was proposed by Dalitz \cite{Dalitz} to be an s-wave NKbar molecule.
And similarly the scalar meson $a_0(980)$ and the $f_0(980)$ meson 
that lie near the KKbar threshold at 995.4 MeV, were proposed by 
Weinstein and Isgur \cite{Wei82} to have the structure of a KKbar 
molecule, and a signature for such a molecular stucture was suggested 
\cite{Clo93} to be given by the ratio of the branching ratios of the 
radiative decay of the $\phi$ meson to the $a_0$ and the $f_0$ 
mesons.

\section{The Molecular Degree of Freedom and Molecular Sum Rules}

A molecular degree of freedom is characterized by excitations that 
involves the relative motion of two tightly bound objects and not 
the excitation of the objects themselve. Hence it is associated 
with a polarization vector known as the separation vector. Such a 
vector can be described geometrically in three dimensions or by using 
the group U(4) \cite{Ia82} and the very succesful Vibron model of 
molecular Physics \cite{Levine}. This model has two symmetry limits that 
corresponds to the geometrical description of Rigid Molecules, the 
O(4) limit, or Soft Molecules, the U(3) limit.

The polarization phenomena associated with a molecular state 
implies that it should be associated with dipole excitations of the 
separation vector. In this case expectation values of the dipole 
operator do not vanish as the center of mass and center of charge 
of the polarized molecular state do not coincide. Hence  
molecular states give rise to low lying dipole excitations. While 
the high lying Giant Dipole Resonace (GDR) is associated with (a 
Goldhaber-Teller) excitation of the neutron distribution against 
that of the proton, a molecular excitation involves a smaller fraction 
of the nucleons at the surface and is thus expected to occur at lower 
excitation; i.e. a soft dipole mode.

The GDR excitation exhausts the (TRK) Energy Weighted Dipole Sum Rule:
\begin{tabbing}
 \hspace{0.5in} $S_1(E1;A)$ \= 
    = $\Sigma_i \ B(E1:0^+ \ \rightarrow \ 1^-_i) \times E^*(1^-_i)$ \\
        \> = ${9 \over 4 \pi} \ {NZ \over A} \times {e^2 \hbar ^2 \over 2m}$
       \hspace{1.5in} \= (equ. 3) \\
\end{tabbing}
And for a molecular state Alhassid, Gai and Bertsch \cite{AGB} 
derived sum rules by subtracting the individual 
sum rules of the contituents from the total sum rule:

\begin{tabbing}
\hspace{0.1in} $S_1(E1;A_1+A_2)$ \= 
   = $S_1(A) \ - \ S_1(A_1) \ - \ S_1(A_2)$ \\
\> = ${{Z_1A_2 \ - A_1Z_2} \over AA_1A_2} \times {e^2 \hbar ^2 \over 2m}$
    \hspace{1.2in} \= (equ. 4) \\
\   \\
\hspace{0.1in} $S_1(E1;\alpha+A_2)$ \> = ${(N-Z)^2 \over A(A-4)} 
 \times {e^2 \hbar ^2 \over 2m} $ \> \\
\hspace{0.1in} $S_1(E1;n+A_2)$ \> = ${Z^2 \over A(A-1)} 
 \times {e^2 \hbar ^2 \over 2m} $ \> \\
\hspace{0.1in} $S_1(E1;2n+A_2)$ \> = ${2Z^2 \over A(A-2)} 
 \times {e^2 \hbar ^2 \over 2m}$ 
\> (equ. 5) 
\end{tabbing}
Note that the sum rule for two neutrons molecular states, 
$S_1(2n+A_2)$, is the same whether one assumes a "di-atomic" 
nuclear molecule ($^9Li$ + a dineutron), or a "tri-atomic" nuclear 
molecule ($^9Li$ + n + n). And the sum rule (as a kinematical 
model) does not allow us to 
distinguish between the two molecular cases.
These molecular sum rules (equs. 4,5) were shown to be useful in elucidating 
molecular (cluster) states in $^{18}O$ where the measured B(E1)'s and 
B(E2)'s exhaust 13\% and 23\%, respectively, of the molecular sum rule
\cite{Ga83}. Similarily, these molecular states in $^{18}O$ 
have alpha widths that exhaust 20\% of the Wigner sum rule. The 
branching ratios for electromagnetic decays in $^{18}O$ were also 
shown to be consistent with predictions of the 
Vibron model in the U(3) limit \cite{Ga91}. Indeed the manifestation 
of a molecular structure in $^{18}O$ has altered our 
undertsanding of the coexistence of degrees of freedoms 
in $^{18}O$ \cite{Ga89}.

The dipole strength at approximately 1.2 MeV in \li11,  
shown in Fig. 1, exhausts 20\% of the two neutrons 
molecular sum rule, and the total 
strength integrated up to 5 MeV exhausts 100\% of that sum rule.
We emphasize that the experimental efficiency at for example 6.0 
MeV is very large (30\%) \cite{GSI}, but no 
strength is found at higher energies 
beyond 100\% of the molecular sum rule. These two facts strongly suggest 
the existence of a low lying soft dipole mode in \li11.

\section{Conclusion}

In conclusions we demonstrate that quantitative tools exist to test 
the validity of the "threshold effect" and the "soft dipole mode" 
interpretation of the dipole (E1) strength in \li11. More precise data  
are needed to rule out one or the other interpretation and this paper 
may serve as an impetus for such data. Current interpration is 
consistent with the existence of a low lying dipole mode in \li11 
at approximately 1.2 MeV, and may pose difficulties to other 
interpretations.


\begin{thebibliography}{99}

\bibitem{GSI} M. Zinser {\em et al.}; Nucl. Phys. {\bf A619}(1997)151.

\bibitem{Ik92} K. Ikeda Nucl. Phys. {\bf A538}(1992)355c.

\bibitem{Ha95} P.G. Hansen; Nucl. Phys. {\bf A588}(1995)1c. \\ P.G. Hansen 
    and A.S. Jensen; Annu. Rev. Nucl. Part. Sci. {\bf 45}(1995)591.

\bibitem{Ko97} A.A. Korsheninnikov {\em et al.}; Phys. 
   Rev. Lett. {\bf 78}(1997)2317.

\bibitem{Kob92} T. Kobayashi; Phys. Lett. {\bf A538}(1992)343c.

\bibitem{Gor98} M.G. Gornov {\em et al.}; Phys. Rev. 
   Lett. {\bf 81}(1998)4325.

\bibitem{Ka97} S. Karataglidis {em et al.}; Phys. Rev. Lett.
    {\bf 79}(1997)1447.

\bibitem{Blatt} J.M. Blatt and V.F. Weisskopf; Theoretical Nuclear 
     Physics, Wiley, 1952.

\bibitem{Sh58} F.L. Shapiro; JETP {\bf 7}(1958)1132.

\bibitem{Ber61} A.A. Bergman, F.L. Shapiro; Sov. Jour. Phys.; 
     JETP {\bf 13}(1961)895.

\bibitem{Bor82} S.B. Borazkov {\em et al.}; Sov. Jour. Nucl. Phys. 
    {\bf 35}(1982)307.

\bibitem{Bethe} H.A. Bethe and C. Longmire; Phys. Rev. 
     {\bf 77}(1950)647.

\bibitem{He98} M. Heil, F. Kappler, M. Wiescher, A. Mengoni; 
     APJ {\bf 507}(1998)997, and references therein.

\bibitem{Levine} F. Iachello and R.D. Levine, Algebraic Theory 
   of Molecules; Oxford University Press, 1995.

\bibitem{Dalitz} R.H. Dalitz; Oxford, Low and Intermediate 
   Energy Kaon-Nucleus Physics, 1981, p. 381.

\bibitem{Wei82} J. Weinstein and N. Isgur; Phys. Rev. Lett. 
   {\bf 48}(1982)659.

\bibitem{Clo93} F. Close, N. Isgur, and S. Kumano; Nucl. 
   Phys. {\bf B389}(1993)513.

\bibitem{Ia82} F. Iachello, and A.D. Jackson; 
     Phys. Lett. {\bf 108B}(1982)151.

\bibitem{AGB} Y. Alhassid, M. Gai, and G.F. Bertsch ; 
     Phys. Rev. Lett. {\bf 49}(1982)1482.

\bibitem{Ga83} M. Gai, M. Ruscev, A.C. Hayes, J.F. Ennis, R. Keddy, 
   E.C. Schloemer, S.M. Sterbenz and D.A. Bromley; 
   Phys. Rev. Lett. {\bf 50}(1983)239.

\bibitem{Ga91} M. Gai {\em et al.}; Phys. Rev. {\bf C43}(1991)2127.

\bibitem{Ga89} M. Gai {\em et al.}; Phys. Rev. Lett. {\bf 62}(1989)874.

\end{thebibliography}
\end{document}